\documentclass[12pt,a4paper]{article}
\usepackage{epsfig}

\newcommand{\<}{\langle}
\renewcommand{\>}{\rangle}

\newcommand{\beq}{\begin{equation}}
\newcommand{\eeq}{\end{equation}}
\newcommand{\beqn}{\begin{eqnarray}}
\newcommand{\eeqn}{\end{eqnarray}}
\newcommand{\nn}{\nonumber}

\usepackage{chir_layout}
\begin{document} 
\begin{titlepage}
\date{\today}
\vspace{-2.0cm}
\title{
{\vspace*{-6mm} 
\begin{flushright}
\small{DESY 05-130}\\
\small{ROM2F/2005/17}\\ 
\small{Bicocca FT-05-17}
\end{flushright}}
\vskip .5cm
\centerline{\LARGE{\bf Chirally improving Wilson fermions}}
\smallskip\centerline{\LARGE{\bf III. The Schr\"odinger functional}}
\vspace*{0mm}}
\author{R.\ Frezzotti$^{a}$\quad
        G.C.\ Rossi$^{b,c}$
        \\[0mm]
  {\small $^a$ INFN, Sezione di Milano} \\
  {\small and Dipartimento di Fisica, Universit\`a di Milano 
   ``{\it Bicocca}''}\\
  {\small Piazza della Scienza 3 - 20126 Milano, Italy}\\
  {\small $^b$ Dipartimento di Fisica, Universit\`a di  Roma
   ``{\it Tor Vergata}''} \\
  {\small and INFN, Sezione di Roma 2}\\
  {\small Via della Ricerca Scientifica - 00133 Roma, Italy}\\
  {\small $^c$ John von Neumann-Institut f\"ur Computing  NIC}\\
  {\small Platanenallee 6, D-15738 Zeuthen, Germany}}
\maketitle
\vspace*{0mm}
\abstract
{We show that it is possible to construct a lattice Schr\"odinger functional
for standard Wilson fermions, where the expectation values of 
${\cal R}_5$-even operators are O($a$) improved, up to terms coming from 
the boundaries.}
\end{titlepage}
\vfill
\newpage

\section{Introduction}
\label{sec:INTRO}

It has been shown in ref.~\cite{FR} that it is possible to improve 
the approach to the continuum limit of correlation functions
in lattice QCD with standard Wilson fermions by taking arithmetic averages 
(Wilson averages -- WA's) of vacuum expectation values (v.e.v.'s) computed 
in theories regularized with opposite values of the Wilson parameter, $r$. 
Improved energies and matrix elements can be obtained by similarly averaging 
the corresponding physical quantities separately computed within the two 
regularizations. The same result can be obtained by replacing the WA 
with a linear combination of v.e.s.'s computed with the same value of $r$, 
but opposite values of the quark mass, $m_q$ (mass average -- MA). 
In our notations $m_q=M_0-M_{\rm cr}$, with $M_0$ the bare quark mass 
parameter and $M_{\rm cr}$ the critical mass. The relevant coefficient
in the linear combination of the MA is the ${\cal R}_5$ parity (see 
eq.~(\ref{PSIBAR}) below) of the operator, $O$, whose v.e.v.\ one 
is computing. Notice that in taking the limit $m_q\to 0$ the two terms 
of the MA may not be equal if spontaneous chiral symmetry breaking occurs.   

To deal with the  problems related to the spectrum of the Wilson--Dirac 
operator~\cite{EXCONF}, twisted-mass lattice QCD (tm-LQCD)~\cite{TM} 
should be better used for actual computations. 
The choice $\omega=\pm \pi/2$ for the twisting angle 
(maximal twist) is particularly interesting, as O($a$) improved estimates 
of all interesting physical quantities can be obtained even without averaging 
data from lattice formulations with opposite Wilson terms. 

These results are valid both in the quenched approximation and in the 
full theory. Indeed it was also shown in ref.~\cite{FRC} how one has to deal 
with the case of mass non-degenerate quark pairs to get O($a$) improvement and 
at the same time a real and positive quark determinant.

Maximally twisted tm-LQCD can also be shown to be a sufficiently flexible
regularization scheme to allow neat solutions~\cite{FRTWO,SINTVLA} of the 
difficult problem that goes under the name of ``wrong chirality 
mixing''~\cite{BMMRT}, which has up to now prevented a (reliable) lattice 
evaluation of the $\Delta I=1/2$ non-leptonic kaon decay amplitudes, if 
Wilson fermions are employed. By coupling the strategy proposed in 
ref.~\cite{FRC} for the sea quark regularization with suitable ''twistings'' 
of the Wilson terms of the valence quarks of various flavours, one can 
rigorously prove that it is possible to get rid of all ``wrong chirality 
mixings'' in the evaluation of the matrix elements of the CP-conserving 
$\Delta S=1,2$ effective weak Hamiltonian.

Striking confirmation of the viability of the approach outlined above and of 
the remarkable properties of tm-LQCD at $|\omega|=\pi/2$ has come from the 
recent works of refs.~\cite{KSUW,XLF,CAN}, where (quenched) studies of the 
scaling behaviour of the theory were carried out down to rather small values 
of lattice spacing and pion mass. Indeed, lattice results for pion masses as 
low as $\sim 250$~MeV show surprisingly small cutoff effects from 
$a\simeq .12$~fm to $a\simeq .05$~fm, if the optimal value of the critical 
mass~\cite{SWNEW,AB,FMPR,PJ,CAN} is employed. A careful study of the 
non-trivial phase structure of the unquenched~\cite{AO,SHASIN,MUSCO} theory 
was also carried out in refs.~\cite{KETAL,MONT} with various choices of the 
gauge action (standard plaquette, DWB2~\cite{DWB2} and tree-level Symanzik 
improved).

A complementary role has been played by the Schr\"odinger functional 
formulation~\cite{SCH,SINT} which has proved to be an invaluable tool in 
applications, especially because it provides a workable scheme for the 
computation of the running of the gauge coupling constant and 
the quark masses~\cite{SCH,QMR}, as well as for the non-perturbative 
evaluation 
of renormalization constants~\cite{RC}. Thus a natural question to ask 
is whether the sort of O($a$) improvement one could get in the infinite 
volume formulation of lattice QCD can also be obtained in the 
Schr\"odinger functional framework.

In this paper we construct a modified Schr\"odinger functional 
describing the gauge interactions of 
a flavour doublet of massless standard Wilson quarks, in which fermions are 
endowed with a sort of twisted boundary conditions. We shall show that, 
under the assumption that IR effects are screened by the natural cutoff 
provided by the finite extension of the time coordinate and up to O($a$) 
(fermionic and gluonic) contributions coming from the boundaries, the 
arithmetic average of the expectation values of ${\cal R}_5$-even 
(multi-local, gauge invariant, multiplicative renormalizable -- m.r.) 
operators, $O$, computed with opposite values of the Wilson parameter 
(WA's) are O($a$) improved. Like in the infinite volume case, the symmetries 
of the theory allow to conclude that the WA is indeed unnecessary, because 
in the massless limit O($a$) bulk lattice artifacts are actually absent.

As for the O($a$) boundary terms, we will see that there are 
two types of them: those coming from the action~\cite{LSSW} and those coming 
from the need of improving operators inserted at the boundaries, if there 
is any such local factor in $O$. 
Boundary terms of the first type will appear in the action with coefficients 
that are even in the Wilson parameter, $r$. Their non-perturbative 
value is not known. In the standard formulation of the Schr\"odinger 
functional given by the $Alpha$-Collaboration they have been computed in 
perturbation theory up to two-loops~\cite{SCH,PT}. It turns out that this 
amount of knowledge is numerically adequate for applications.  
Boundary terms of the second type appear multiplied by 
coefficients that do not have definite $r$-parity, owing to the breaking 
of parity (see eq.~(\ref{PAROP}) induced by our choice of fermionic 
boundary conditions (see below).

As we said, in this paper we will limit ourselves to discuss the massless
theory, which is enough for the non-perturbative evaluation of fermionic 
renormalization constants and the computation of the running of the gauge 
coupling and quark masses. 

\section{The construction of the Schr\"odinger functional}
\label{sec:SFNEW}

The problem with exporting the philosophy of the approach of ref.~\cite{FR} 
to the standard Schr\"odinger functional formulation~\cite{SCH,SINT} 
is related to the fact that, when the Schr\"odinger functional is 
constructed via the iteration of the transfer matrix 
operator~\cite{ML}~\footnote{See ref.~\cite{PREV} for other attempts to 
define the Schr\"odinger functional in the continuum and on the lattice.}, 
the projectors that define which fermionic components have to be 
prescribed at the boundaries are completely determined by the form of 
the transfer matrix and ultimately of the lattice action. For instance, 
for Wilson fermions the boundary conditions must have the Dirichlet form 
\beqn
&P_{+}\psi({\bf x},0)=\rho({\bf x})\, ,\quad
&P_{-}\psi({\bf x},T)=\rho'({\bf x})\, ,\label{BCOLD}\\
&\bar\psi({\bf x},0)P_{-}=\bar\rho({\bf x})\, ,\quad
&\bar\psi({\bf x},T)P_{+}=\bar\rho'({\bf x})\, ,\nn\eeqn
with
\beq
P_{\pm}=\frac{1}{2}(1\pm r\gamma_0)\, .\label{PROLD}\eeq
The choice $|r|=1$ of the Wilson parameter is compulsory, in order for
spin operators, $P_\pm$, to be true projectors ($P_\pm^{\,2}=P_\pm$).
 
Despite the fact that the finite time lattice action defined in this 
way can be shown to enjoy the (spurionic) symmetries that in the 
infinite volume theory were sufficient to guarantee O($a$) improvement 
of Wilson averages~\cite{FR}, a similar conclusion cannot be 
drawn here, because upon inverting the sign of $r$ the expression of the 
Schr\"odinger functional will change by O(1) terms. This operation, in fact,
also affects the form of the boundary conditions since $P_\pm\to P_\mp$ 
under $r\to -r$.

The only envisageable way out of this difficulty is to have a formulation 
where the structure of the Wilson term and the form of the fermionic 
boundary conditions are not correlated~\footnote{Though developed with the 
purpose of constructing a lattice Schr\"odinger functional for overlap 
fermions, the interesting orbifold construction of ref.~\cite{TAN} was of 
some inspiration to us. Ideas similar to the ones discussed here were also 
presented in ref.~\cite{SINTVM}.}. We then propose to change the spin 
projectors from~(\ref{PROLD}) and consider a situation where 
homogeneous ${\cal R}_5$-invariant constraints are taken. 
More precisely, we propose to construct a Schr\"odinger-like functional 
where fermions are introduced in pairs and obey the following homogeneous 
boundary conditions 
\beqn
&\Pi_{+}\psi({\bf x},0)=0\, ,\quad
&\Pi_{-}\psi({\bf x},T)=0\, ,\nn\\
&\bar\psi({\bf x},0)\Pi_{-}=0\, ,\quad
&\bar\psi({\bf x},T)\Pi_{+}=0\, ,\label{BCHOM}\eeqn
with
\beq
\Pi_{\pm}=\frac{1}{2}(1\pm \tau_3\gamma_5)\, .\label{PRNEW}\eeq
Although not immediately required for O($a$) improvement, we are 
imagining that quarks are introduced in pairs with opposite chiral 
boundary projectors for the two flavour components (say, up and down). 
As we shall see, this is necessary to have a real and positive determinant.  

With only the modification~(\ref{BCHOM}) of the boundary conditions, the 
improvement of the expectation value of m.r.\ ${\cal R}_5$-even operators 
(up to O($a$) contaminations from boundary lattice artifacts) follows 
from symmetry arguments very similar to those that have been used in the 
infinite volume theory.

We now wish to give a few more details about our construction. 
Let us start by writing down the general form of the action 
of a pair of massless Wilson fermions obeying the boundary 
conditions~(\ref{BCHOM}), extended over the finite time interval $[0,T]$. 
Adapting the analysis of ref.~\cite{SINT} to the present 
situation and separating out the boundary terms from the rest, we write 
\beqn
S_{\rm{F}}=S_{\rm{bulk}}+S_{B}^{i}+S_{B}^{f}\, ,\label{FERACT}\eeqn 
\beqn 
\hspace{-.6cm}&&S_{\rm{bulk}}
=-\frac{a^3}{2}\sum_{\bf x}\sum_{t=a}^{T-2a}\,\Bigl[
\bar\psi({\bf x},t)U_0({\bf x},t)(r-\gamma_0)
\psi({\bf x},t+a)+\nn\\
\hspace{-.6cm}&&\qquad+\bar\psi({\bf x},t+a)(r+\gamma_0)
U_0^\dagger({\bf x},t)\psi({\bf x},t)\Bigr]+\nn\\
\hspace{-.6cm}&&-\frac{a^4}{2a}\sum_{{\bf x},k}\sum_{t=a}^{T-a}\,
\Bigl[\bar\psi({\bf x},t)U_k({\bf x},t)
(r-\gamma_k)\psi({\bf x}+\hat{k},t)+
\nn\\\hspace{-.6cm}&&\qquad+
\bar\psi({\bf x}+\hat{k},t)(r+\gamma_k)U_k^\dagger({\bf x},t)
\psi({\bf x},t)\Bigr]+\nn\\
\hspace{-.6cm}&&+a^4\sum_{\bf x}\sum_{t=a}^{T-a}\bar\psi({\bf x},t)
\Bigl[M_{\rm{cr}}(r)+\frac{4r}{a}\Big{]}\psi({\bf x},t)\label{BULK}\, ,
\eeqn
\beqn
\hspace{-.8cm}&&S_{B}^{i}=-\frac{a^3}{2}\sum_{\bf x}
\Bigl[\Big{(}\bar\psi({\bf x},0)\Pi_{+}U_0({\bf x},0)
(r-\gamma_0)\psi({\bf x},a)+\nn\\
\hspace{-.8cm}&&\qquad\qquad+\bar\psi({\bf x},a)
(r+\gamma_0)U_0^\dagger({\bf x},0)\Pi_{-}\psi({\bf x},0)\Bigr]
+\nn\\\hspace{-.8cm}&&+\frac{a^3}{2}\sum_{{\bf x},k}
\bar\psi({\bf x},0)\Pi_{+}\gamma_k\Bigl[V^i_k({\bf x})\psi({\bf x}+\hat k,0)
-V_k^{i\dagger}({\bf x})
\psi({\bf x}-\hat k,0)\Bigr]\, ,\label{BOUNDI}\eeqn
\beqn
\hspace{-.8cm}&&S_{B}^{f}
=-\frac{a^3}{2}\sum_{\bf x}\Bigl[\bar\psi({\bf x},T-a)
U_0({\bf x},T-a)(r-\gamma_0)\Pi_{+}\psi({\bf x},T)+\nn\\
\hspace{-.8cm}&&\qquad\qquad+\bar\psi({\bf x},T)\Pi_{-}
(r+\gamma_0)U_0^\dagger({\bf x},T-a)\psi({\bf x},T-a)\Bigr]+\nn\\
\hspace{-.8cm}&&+\frac{a^3}{2}\sum_{{\bf x},k}
\bar\psi({\bf x},T)\Pi_{-}\gamma_k\Bigl[V^f_k({\bf x})\psi({\bf x}+\hat k,T)
-V_k^{f\dagger}({\bf x})\psi({\bf x}-\hat k,T)\Bigr]
\, .\label{BOUNDF}\eeqn
For clarity in eqs.~(\ref{BULK}) to~(\ref{BOUNDF}) we have kept 
separated spatial {\it vs}\ time components and sums. As usual we are 
assuming periodicity in space and fix the spatial components of the gauge
links at the two time boundaries through
\beq
U_k({\bf x}, 0) = V^{i}_k({\bf x}) \, , \qquad 
U_k({\bf x}, T) = V^{f}_k({\bf x}) \, , \qquad  k=1,2,3 \, .
\label{UTBC}
\eeq
In eq.~(\ref{BULK}) $M_{\rm cr}(r)$ is the critical fermion mass. We 
remark that no mass terms for the fermion fields living at the boundaries 
have been included. We will clarify the reason for that in the next sections. 

With the action~(\ref{FERACT}) we define a Schr\"odinger functional 
through the formula 
\beq
{\cal{K}}^{\rm W}\Big{[}V^{f};V^{i}\Big{]}\!=\!
\int\!{\cal{D}}\mu_G[U]\!\int\!{\cal{D}}\mu_F[\bar\psi,\psi]
{\rm{e}}^{-S_{{\rm YM}}-S_{\rm{F}}}\, ,\label{KSCH}\eeq
where $S_{\rm YM}$ is the pure gauge finite time action~\cite{SCH} 
The gauge integration measure, ${\cal{D}}\mu_G[U]$, is as explained in 
refs.~\cite{SCH,SINT}. The fermionic integration measure, 
${\cal{D}}\mu_F[\bar\psi,\psi]$, is spatially periodic and it is 
extended in time to all fermionic variables from $t=a$ to $t=T-a$ and 
over the variables $\Pi_{-}\psi({\bf x},0),\bar\psi({\bf x},0)\Pi_{+}$ and 
$\Pi_{+}\psi({\bf x},T),\bar\psi({\bf x},T)\Pi_{-}$ at the initial 
and final time, respectively. 

It should be noted that this kernel enjoys the usual convolution 
properties only in the continuum limit, unlike the situation one has 
in the construction considered in refs.~\cite{SCH,SINT}. Notice that, 
since the kernel~(\ref{KSCH}) is not the iteration of the transfer matrix, 
one could even have dropped the terms proportional to $r$ in 
eqs.~(\ref{BOUNDI}) and~(\ref{BOUNDF}).

\section{Symmetry properties}
\label{sec:SYMPRO}

The key observation of this paper is that the action~(\ref{FERACT}) is 
invariant under a large set of transformations, which leave unaltered 
the structure of the homogeneous fermionic boundary 
constraints~(\ref{BCHOM}). They are 
\begin{itemize}
\item ${\cal{R}}_5\times (r\to -r)$, where 
\begin{equation}
{\cal{R}}_5 : \left \{\begin{array}{ll}
&\hspace{-.3cm}\psi(x)\rightarrow\psi'(x)=\gamma_5 \psi(x)\\
&\hspace{-.3cm}\bar{\psi}(x)\rightarrow\bar{\psi}'(x)=-\bar{\psi}(x)\gamma_5
\end{array}\right . \label{PSIBAR} \quad\eeq
\item ${\cal{R}}_5\times {\cal D}_d \times {\mathcal {P}} 
\times (V^{f}_k \leftrightarrow V^{i}_k) $, where 
($x_D=(-{\bf {x}},T-x_0)$)
\begin{equation}
{\cal{D}}_d : \left \{\begin{array}{lll}    
&\hspace{-.3cm}\psi(x)\rightarrow e^{3i\pi/2} \psi(x_D) \\  
&\hspace{-.3cm}\bar{\psi}(x)\rightarrow e^{3i\pi/2} \bar{\psi}(x_D) \\
&\hspace{-.3cm}U_0(x)\rightarrow U_0^\dagger(x_D-a\hat 0) \\
&\hspace{-.3cm}U_k(x)\rightarrow U_k^\dagger(x_D-a\hat k)\, ,\quad k=1,2,3
\end{array}\right . \label{FIELDTC} \end{equation}
and ${\mathcal {P}}$ is the standard parity operation 
($x_P=(-{\bf {x}},x_0)$)
\beqn
{\mathcal {P}}:\left \{\begin{array}{ll}
&\hspace{-.3cm}\psi(x)\rightarrow \gamma_0 \psi(x_P)\\
&\hspace{-.3cm}\bar{\psi}(x)\rightarrow\bar{\psi}(x_P)\gamma_0\\
&\hspace{-.3cm}U_0(x)\rightarrow U_0(x_P)\, ,\\
&\hspace{-.3cm}U_k(x)\rightarrow U_k^{\dagger}(x_P-a\hat{k})
\, ,\quad k=1,2,3
\end{array}\right . \label{PAROP}\eeqn
\item the product ${\mathcal {C}}{\mathcal {P}}$, where ${\mathcal {C}}$ 
is charge conjugation ($^{\,T}$ means transposition)
\beqn
{\mathcal {C}}:\left \{\begin{array}{ll}
&\hspace{-.3cm}\psi(x)\rightarrow i\gamma_0 \gamma_2\bar\psi(x)^{T}\\
&\hspace{-.3cm}\bar{\psi}(x)\rightarrow -{\psi}(x)^{T}i\gamma_0\gamma_2\\
&\hspace{-.3cm}U_\mu(x)\rightarrow U_\mu^{\star}(x)\, ,\quad \mu=0,1,2,3
\end{array}\right . \label{CHAROP}\eeqn
\item time inversion around the mid-point $T/2$, 
${\cal T}\times (V^{f}_k \leftrightarrow V^{i}_k)$, where ($x_T=({\bf {x}},T-x_0)$)
\beqn
{\mathcal {T}}:\left \{\begin{array}{ll}
&\hspace{-.3cm}\psi(x)\rightarrow \gamma_0 \gamma_5\psi(x_T)\\
&\hspace{-.3cm}\bar{\psi}(x)\rightarrow\bar{\psi}(x_T)\gamma_5\gamma_0\\
&\hspace{-.3cm}U_0(x)\rightarrow U_0^{\dagger}(x_T-a\hat{0})\, ,\\
&\hspace{-.3cm}U_k(x)\rightarrow U_k(x_T)
\, ,\quad k=1,2,3  
\end{array}\right . \label{TINVOP}\eeqn
\item cubic $H(3)$ group
\end{itemize} 
We explicitely notice that the breaking of ${\cal P}$, ${\cal C}$ and 
${\cal{R}}_5\times {\cal D}_d$ is entirely due to the boundary action 
terms~(\ref{BOUNDI}) and~(\ref{BOUNDF}).

Besides the transformations collected above, which are all flavour diagonal, 
we have also invariance under
\begin{itemize}
\item the vector rotation in the iso-spin direction 3
\beqn
{\mathcal {I}}_3:\left \{\begin{array}{ll}
&\hspace{-.3cm}\psi(x)\rightarrow e^{i\omega \tau_3/2}\psi(x)\\
&\hspace{-.3cm}\bar{\psi}(x)\rightarrow {\psi}(x)e^{-i\omega \tau_3/2}\\
\end{array}\right . \label{I3}\eeqn
\item and the four transformations separately acting on only 
the fermionic fields at the boundaries
\beqn
{\mathcal {\cal A}}_{3}^i:
&\hspace{-.3cm}\psi({\bf x},0)\rightarrow -\gamma_5\tau_{3}\psi({\bf x},0)\, ,
\label{PAROP31}\\
\overline{{\mathcal {\cal A}}}_{3}^i:
&\hspace{-.3cm}\bar{\psi}({\bf x},0)\rightarrow
\bar{\psi}({\bf x},0)\gamma_5\tau_{3}\, ,\label{PAROP32}\\
{\mathcal {\cal A}}_{3}^f:
&\hspace{-.3cm}\psi({\bf x},T)\rightarrow\gamma_5\tau_{3}\psi({\bf x},T)\, ,
\label{PAROP33}\\
\overline{{\mathcal {\cal A}}}_{3}^f:
&\hspace{-.3cm}\bar{\psi}({\bf x},T)\rightarrow
-\bar{\psi}({\bf x},T)\gamma_5\tau_{3}\, .\label{PAROP34}\eeqn
\end{itemize}
Finally we mention the reflection symmetry, 
$\Theta\times (V^{f}_k \leftrightarrow V^{i}_k)$, 
related to reflection positivity~(see e.g.~\cite{MON,FR,FRC}).
$\Theta$ is defined to act as follows
\beqn
&&\Theta[f(U)\psi(x_1)\ldots\bar\psi(x_n)]=\nonumber\\
&&=f^\star(\Theta [U])\Theta[\bar\psi(x_n)]\ldots\Theta[\psi(x_1)]\, ,
\label{TMON}\eeqn
where $f(U)$ is a functional of link variables and for $a\leq x_0\leq T-a$
\beqn \left \{\begin{array}{ll}
&\hspace{-.3cm}\Theta_{s/\ell}[\psi(x)]=\bar\psi(\theta_{s/\ell} x)
\gamma_0\nonumber\\
&\hspace{-.3cm}\Theta_{s/\ell}[\bar\psi(x)]=
\gamma_0\psi(\theta_{s/\ell} x)\\
&\hspace{-.3cm}\Theta_{s/\ell}[U_k(x)]=U_k^\star(\theta_{s/\ell} x)\nonumber\\
&\hspace{-.3cm}\Theta_{s/\ell}[U_0(x)]=
U_0^T(\theta_{s/\ell} x - a\hat{0})\label{TUUD}
\end{array}\right . \label{TPPB}\eeqn
with
\beq
\theta_{s/\ell}({\bf{x}},x_0)=({\bf{x}},T-x_0)\label{TETA}
\eeq
to be interpreted as a site ($s$) or link ($\ell$) time-reflection
depending on whether $N_T$ ($T=N_T a$) is odd or even, respectively.
The action of $\Theta$ on the fields at the boundaries is
\beqn \left \{\begin{array}{ll}
&\hspace{-.3cm}\Theta_{s/\ell}\Pi_{-}\psi({\bf x},0)]=
\bar\psi({\bf x},T)\Pi_{-}\gamma_0\nonumber\\
&\hspace{-.3cm}\Theta_{s/\ell}\bar\psi({\bf x},0)]\Pi_{+}=
\gamma_0\Pi_{+}\psi({\bf x},T)\\
&\hspace{-.3cm}\Theta_{s/\ell}\Pi_{+}\psi({\bf x},T)]=
\bar\psi({\bf x},0)\Pi_{+}\gamma_0\nonumber\\
&\hspace{-.3cm}\Theta_{s/\ell}\bar\psi({\bf x},T)]\Pi_{-}=
\gamma_0\Pi_{-}\psi({\bf x},0)
\end{array}\right . \label{BPP}\eeqn

\section{Lattice artifacts}
\label{sec:LA}

Our aim in this section is to analyze the lattice artifacts affecting 
the (on-shell) expectation value
\beq
\<O\>_{T}=
\int\!{\cal{D}}\mu_G[U]\!\int\!{\cal{D}}\mu_F[\bar\psi,\psi] 
O(\psi,\bar\psi,U)\,{\rm{e}}^{-S_{{\rm YM}}-S_{\rm{F}}}
\, ,\label{EXPV}\eeq
where $O$ is a m.r. (multi-local) operator.

The analysis will be based on the assumption that the approach to the 
continuum limit of the Schr\"odinger functional defined in eq.~(\ref{KSCH}) 
can be described by a local effective theory. The latter will be 
characterized a local effective Lagrangian (LEL) including bulk and boundary 
terms~\cite{LSSW}. In determining the operators 
contributing to the Symanzik description of the lattice expectation 
value~(\ref{EXPV}) a key role will be naturally played by the symmetries 
collected in sect.~\ref{sec:SYMPRO}. Let us now go through 
the list of the allowed operators in the order of increasing dimension. 

$\bullet$ The basic observation is that no dimension 3 operators can be 
generated through radiative corrections in the boundary LEL, because operators 
of the form $\bar\psi\Gamma(1\!\!1/\tau_b)\psi$ (with $\Gamma$ any of the 16 
independent Dirac matrix and with or without the insertion of an iso-spin 
matrix) are all forbidden by some of the symmetries above.

$\bullet$ At dimension 4 there is a number of contributions to 
the boundary LEL. Besides the terms corresponding to the continuum 
(infinite volume) QCD action, there will be O($a$) boundary 
terms of the form (see eqs.~(\ref{BOUNDI}) and~(\ref{BOUNDF}))
\beqn
\int d{\bf x}\,\bar\psi({\bf x},0)\Pi_{+}\gamma_k D_k\psi({\bf x},0)
\, ,\quad
\int d{\bf x}\,\bar\psi({\bf x},T)\Pi_{-}\gamma_k D_k\psi({\bf x},T)
\, .\label{OABT}\eeqn
The temporal analogs of these terms can be ignored in the present analysis, 
as they can always be traded for the previous ones using the field 
equations of motion. Because of the symmetry ${\cal{R}}_5\times (r\to -r)$ 
the operators~(\ref{OABT}) being even under ${\cal R}_5$ will intervene 
multiplied by coefficients that will be even functions of $r$. 

There will also be the O($a$) pure gauge boundary terms 
\beqn
&&\int d{\bf x}\,{\rm tr}(F_{0k}F_{0k})({\bf x},0)\, ,\quad 
\int d{\bf x}\,{\rm tr}(F_{jk}F_{jk})({\bf x},0)\, , \nn\\
&&\int d{\bf x}\,{\rm tr}(F_{0k}F_{0k})({\bf x},T)\, ,\quad 
\int d{\bf x}\,{\rm tr}(F_{jk}F_{jk})({\bf x},T)\, ,\label{CTG}
\eeqn
always with coefficients even in $r$, while the $\tilde FF$ term is 
excluded by ${\cal CP}$.

Notice that, if in the multi-local operator, $O$, there are boundary local 
factors, i.e.\ factors, $O_B$, which in the continuum limit are kept at 
vanishing distance from the boundaries, there may be (if not excluded 
by the symmetries listed in sect.~\ref{sec:SYMPRO}) O($a$) terms coming 
from the collision of $O_B$ with the spatially integrated boundary terms of
eqs.~(\ref{OABT}) and~(\ref{CTG}) above.

$\bullet$ At dimension 5 there are O($a$) effects in the expectation 
value $\<O\>_{T}$ that in the Symanzik language are described  by the 
insertion, together with $O$, of the integrated (dimension 5 and parity even) 
bulk LEL density ${\cal L}_5$. Besides these terms there will also appear bulk 
contact terms, coming from the collision under integration of ${\cal L}_5$ 
with the various local bulk factors of $O$. 
In opposition with the previously mentioned local boundary factors,
local bulk factors are (products of local) operators that in the continuum 
limit are kept at finite distance in physical units from the boundaries.

Because of the symmetries ${\cal{R}}_5\times (r\to -r)$ and 
${\cal{R}}_5\times {\cal D}_d \times {\mathcal {P}} \times {\mathcal {T}}$, all 
such O($a$) bulk contributions will appear with coefficients odd in $r$. 
This conclusion follows also here thanks to the fact that 
in the continuum LEL parity is only broken by boundary 
terms which are not relevant for bulk operators, while the bulk LEL 
density ${\cal L}_5$ is still even under parity and
${\mathcal {T}}$ and consequently odd under ${\cal{R}}_5$. 
Thus just like in the infinite volume formulation, 
all O($a$) bulk contributions in the Symanzik expansion will cancel out 
if the expectation value of $O$ computed with $r$ is averaged with 
its expectation value computed with $-r$, i.e.\ in the quantity 
\beq
\<O\>_{T}\Big{|}_{\rm WA}=
\frac{1}{2}\Big{[}\<O\>_{T}(r)+\<O\>_{T}(-r)\Big{]}
\, .\label{EXPWA}\eeq
Notice that in order to compute $\<O\>_{T}|_{\rm WA}$ it is not necessary 
to perform two independent simulations with opposite value of $r$. It is 
enough to notice the relation
\beq
\<O\>_{T}(-r)=(-1)^{{\cal P}_5[O]}\<O\>_{T}(r)
\, ,\label{EXPMR}\eeq
where ${{\cal P}_5[O]}$ is the parity of $O$ under the transformation 
${\cal R}_5$. Introducing eq.~(\ref{EXPMR}) in~(\ref{EXPWA}), we 
conclude that for ${\cal R}_5$-even operators O($a$) bulk terms are 
actually absent in the chiral limit and this is why only one simulation 
is needed to get O($a$) bulk improvement. If instead $O$ is an 
${\cal R}_5$-odd operator, an identically vanishing result is obtained
from the WA. In the absence of spontaneous symmetry breaking phenomena 
affecting the expectation value of $O$, this is indeed the result we 
would like to get in the continuum. In fact $\<O\>_{T}(r)$ 
is itself an O($a$) quantity.  

\subsection{Boundary operators and O($a$) artifacts: an example}
\label{sec:AIE}

As an example, we briefly discuss a possible setting 
for the non-perturbative computation of the renormalization constant, $Z_P$, 
of the pseudoscalar density operator, $\bar\psi\gamma_5\tau_b\psi$, 
where $b=1,2,3$ is an isospin index. The knowledge of $Z_P$ is relevant, 
for instance, for the study of the running of the quark mass~\cite{QMR}.

To extract $Z_P$ from simulation data it is sufficient to compute 
the following lattice expectation values (for definiteness we set the 
isospin index $b$ equal to 1)
\beq
\<\,(\bar\psi\gamma_5\tau_1\psi)({\bf 0},T/2) \Phi^i_{P1S2}\,\>_{T}
\qquad {\rm and}\quad
\<\,\Phi^f_{P1S2}\,\Phi^i_{P1S2}\,\>_{T}\, ,\label{ZP}\eeq 
where $\Phi^i_{P1S2}$ and $\Phi^f_{P1S2}$ are zero three-momentum operators 
sitting at the initial ($x_0=0$) and final ($x_0=T$) times, respectively, 
given by
\beqn
\Phi^i_{P1S2}= a^3\sum_{\bf x}\bar\psi({\bf x},0)\Pi_+ 
\frac{1}{2} (\gamma_5\tau_1 + i\tau_2) \Pi_- \psi({\bf x},0) \, , 
\nonumber \\
\Phi^f_{P1S2}= a^3\sum_{\bf x}\bar\psi({\bf x},T)\Pi_-
\frac{1}{2} (\gamma_5\tau_1 - i\tau_2) \Pi_+ \psi({\bf x},T) \,
\, .\label{BZP}
\eeqn
Notice that $\Phi^i_{P1S2}$ and $\Phi^f_{P1S2}$ are defined in terms of the 
dynamical boundary quark components only. The presence of the projectors 
$\Pi_\pm$ implies that the Dirac-flavour structure 
necessarily appears in the combinations $(\gamma_5\tau_1 + i\tau_2)$ and 
$(\gamma_5\tau_1 - i\tau_2)$ at $x_0=0$ and $x_0=T$, respectively.

An analysis of the dimension three and four boundary operators compatible 
with the symmetries listed in sect.~\ref{sec:SYMPRO} shows that
the fields~(\ref{BZP}) renormalize multiplicatively and, once summed over 
their spatial argument, are free from O($a$) corrections. 

Putting this result together with that on the O($a$) bulk improvement
reached in sect.~\ref{sec:LA}, one arrives at the conclusion that the 
calculation of $Z_P$ carried out with the Schr\"odiger functional formalism 
developed in this paper is free from O($a$) discretization errors, 
except of course for those coming from the boundary gauge link 
operators in the first column of eq.~(\ref{CTG}), which survive even 
at vanishing values of the bounday gauge links. 

Although the Schr\"odinger functional setup proposed 
in this paper allows more freedom (than apparent from the 
above example) in building fermionic boundary operators~\footnote{This 
interesting matter is left for a future study.}, the important point
we would like to make here is that in several cases it may be possible
to choose the boundary operators in $O$ (those denoted above as $O_B$) in
such a way that $\< O \>_{T}$ is free from O($a$) cutoff effects, once
the lattice action has been supplemented with boundary counterterms.

\section{Fermion determinant}
\label{sec:DET}

In this section we want to show that, associated with the fermionic 
integration defined in eq.~(\ref{KSCH}), there is a real and positive 
determinant. This feature is necessary for the proper interpretation 
of the fermionic contribution to the functional integral 
as a well defined weight for the successive gauge integration and the 
viability of simulations. It is precisely the need to fulfill this 
requirement that has led us to introduce a flavour doublet of fermions 
endowed with opposite chirality boundary conditions

The situation for the fermion integration is very similar to the one 
discussed in sect.~6 of ref.~\cite{SINT}. Calling $u$ and $d$ the two 
flavour components of $\psi$, we can define the two pre-Hilbert spaces 
${\cal H}_{u}$ and ${\cal H}_{d}$ as the spaces of spinors satisfying the 
conditions dictated by the eqs.~(\ref{BCHOM}), namely  
\beqn
&&{\cal H}_{u}=\{u\,|\,(1+\gamma_5)u({\bf x},0)=0,
\,(1-\gamma_5)u({\bf x},T)=0\}\, ,\label{HPU}\\
&&{\cal H}_{d}=\{d\,|\,(1-\gamma_5)d({\bf x},0)=0,
\,(1+\gamma_5)d({\bf x},T)=0\}\, .\label{HPD}\eeqn
With this splitting the action~(\ref{FERACT}) can be rewritten  
in terms of the operator 
\beq {\cal D}=\left(\begin{array}{cc}
0&D_u\\D_d&0 \end{array}\right)\label{MAT}\eeq
through the formula~\footnote{Not to obscure the argument we have 
suppressed space-time indices.}
\beq
S_{\rm F}=\left(\begin{array}{cc}\bar u & \bar d\end{array}\right) 
\left(\begin{array}{cc} 0&D_u\\D_d&0\end{array}\right)
\left(\begin{array}{c} d \\ u \end{array}\right)=
\bar u D_{u} u +\bar d D_{d} d \, .\label{ACTMAT}\eeq
One notices that the operator ${\cal D}$ admits a well defined 
eigenvalue problem in the Hilbert space ${\cal H}_{u}\oplus{\cal H}_{d}$,
with the functional integration over the fermionic variables $u$ and $d$ 
giving as a result precisely the determinant of $D$.

Furthermore we observe that for the purpose of computing the fermionic 
functional integral in~(\ref{KSCH}) we can replace the operator~(\ref{MAT})
with 
\beq \widetilde{{\cal D}} =\left(\begin{array}{cc}
0&D_u\\\gamma_5D_d\gamma_5&0 \end{array}\right)\label{MAT5}\, .\eeq
This replacement only amounts to the harmless change of integration variables
\beq d\to\gamma_5 d\, \qquad \bar d\to \bar d \gamma_5\, .\label{CHINT}\eeq
The reason for doing so is that in this way one gets an operator, 
$\widetilde{{\cal D}}$, which is self-adjoint since 
\beq \gamma_5D_d\gamma_5=D_u^\dagger\, ,\label{HERM}\eeq
as one can explicitly check with some algebra. Consequently the 
determinant of $\widetilde{{\cal D}}$ will be a real number. 
Actually this number, hence the determinant of ${\cal D}$, is also positive.

To prove this statement it is convenient to consider the 
auxiliary operator
\beq {\cal M}=\left(\begin{array}{cc}
0&\Gamma_0\\\Gamma_0&0\end{array}\right)\left(\begin{array}{cc}
0&D_u\\D_u^\dagger&0 \end{array}\right)=\left(\begin{array}{cc}
\Gamma_0D_u^\dagger&0\\0&\Gamma_0D_u \end{array}\right)\label{AUX}\, ,\eeq
where $\Gamma_0$ is a block diagonal matrix in the time direction 
with the following structure. All along the diagonal 
there are $N_T-1$ copies (recall $T=N_T a$) of the $4\times 4$ usual 
$\gamma_0$ Dirac-matrix, except in the upper-left and the lower-right 
corners where there is a $2\times 2$ unit matrix. One checks that ${\cal M}$ 
maps ${\cal H}$ into itself, because 
\beqn
\Gamma_0 D_u &:& {\cal H}_{u}\to {\cal H}_{u}\, ,\nn\\
\Gamma_0 D_u^\dagger &:& {\cal H}_{d}\to {\cal H}_{d}\, .\label{CORR}
\eeqn
Positivity of ${\rm det}[D]$ then follows from the chain 
of equalities 
\beq {\rm det}[D]= {\rm det}[\widetilde{\cal D}]={\rm det}[{\cal M}]=
{\rm det}[D_u^\dagger\Gamma_0]\,{\rm det}[\Gamma_0D_u]=
|{\rm det}[\Gamma_0D_u]|^2\, ,\label{POUD}\eeq
in which we have used the observation that the determinant of the 
off-diagonal matrix where $\Gamma_0$ intervene in eq.~(\ref{AUX}) 
is equal to 1.

\section{Conclusions}
\label{sec:CONC}

We have shown in this paper that it is possible to modify the usual 
definition of the Schr\"odinger functional in a way that allows to 
compute expectation values of ${\cal R}_5$-even (multi-local) operators 
having O($a$) cutoff effects only coming from the boundaries. Assuming that 
the latter are under control, either because they are just absent, owing to
symmetry reasons, or because known to some order in perturbation theory, 
this approach provides a viable scheme for the improved computation of 
the renormalization constants of fermionic operators and the running 
of QCD parameters. 

\vspace{0.5cm}
\noindent{\bf Acknowledgments} - One of us (G.C.R.) is grateful to the 
Humboldt Foundation for partial financial support. Partial supported 
from M.I.U.R. (Italy) is also acknowledged.


\begin{thebibliography}{99}

\bibitem{FR}
R. Frezzotti and G.C. Rossi, JHEP {\bf 0408} (2004) 007 (hep-lat/0306014) and 
Nucl. Phys. {\bf B} (Proc. Suppl.) {\bf 129} (2004) 880 (hep-lat/0309157).
%hep-lat/0306014 and hep-lat/0309157.

\bibitem{EXCONF}
W.A. Bardeen, A. Duncan, E. Eichten, G. Hockney and H. Thacker,
%``Light quarks, zero modes, and exceptional configurations,''
Phys. Rev.  {\bf D57} (1998) 1633;\\
W.A. Bardeen, A. Duncan, E. Eichten and H. Thacker,
%``Quenched chiral artifacts for Wilson-Dirac fermions,''
Phys. Rev. {\bf D59} (1999) 014507;\\
G. Schierholz et al., Nucl. Phys. {\bf B} (Proc. Suppl.) {\bf 73} (1999) 889.

\bibitem{TM} 
R. Frezzotti, P.A. Grassi, S. Sint and P. Weisz, Nucl. Phys. {\bf B} (Proc.
Suppl.) {\bf 83} (2000) 941 and JHEP {\bf 0108} (2001) 058;\\
R. Frezzotti, S. Sint and P. Weisz [ALPHA Collaboration], JHEP {\bf 0107}
(2001) 048 (hep-lat/0104014);\\
M. Della Morte, R. Frezzotti, J. Heitger and S. Sint [ALPHA Collaboration],
JHEP {\bf 0110} (2001) 041;\\
R. Frezzotti and S. Sint, Nucl. Phys. {\bf B} (Proc. Suppl.) {\bf 106}
(2002) 814;\\
R. Frezzotti, Nucl. Phys. {\bf B} (Proc. Suppl.) {\bf 119} (2003) 140 and
Nucl. Phys. {\bf B} (Proc. Suppl.) {\bf 140} (2005) 134, hep-lat/0409138.
%R. Frezzotti, P.A. Grassi, S. Sint and P. Weisz, Nucl. Phys. {\bf B} (Proc.
%Suppl.) {\bf 83} (2000) 941 and JHEP {\bf 0108} (2001) 058;\\
%R. Frezzotti, Nucl. Phys. {\bf B} (Proc. Suppl.) 
%{\bf 119} (2003) 140, hep-lat/0210007;\\
%R. Frezzotti, S. Sint and P. Weisz, JHEP {\bf 0107} (2001) 048;\\
%M. Della Morte, R. Frezzotti, J. Heitger and S. Sint  [ALPHA
%collaboration], JHEP {\bf 0110} (2001) 041;\\
%R. Frezzotti and S. Sint, Nucl. Phys. {\bf B} (Proc. Suppl.) {\bf 106}
%(2002) 814, hep-lat/0110140.

\bibitem{FRC}
R. Frezzotti and G.C. Rossi,  Nucl. Phys. {\bf B} (Proc. Suppl.)
{\bf 128} (2004) 193 (hep-lat/0311008).

\bibitem{FRTWO}
R. Frezzotti and G.C. Rossi, JHEP {\bf 0410} (2004) 070, hep-lat/0407002.

\bibitem{SINTVLA}
%C. Pena, S. Sint, and A. Vladikas, hep-lat/0209045.
C. Pena, S. Sint and A. Vladikas, JHEP {\bf 0409} (2004) 069, 
hep-lat/0405028;\\
%``Twisted mass QCD and lattice approaches to the $\Delta I = 1/2$ rule,''
%%CITATION = HEP-LAT 0405028;%%
%``Towards a determination of g(8) and g(27) from twisted mass lattice  QCD,''
%%CITATION = NUPHZ,129,263;%%
%``Twisted mass QCD and the Delta(I) = 1/2 rule,''
Nucl. Phys. {\bf B} (Proc. Suppl.)  {\bf 119} (2003) 368, hep-lat/0209045;
%%CITATION = HEP-LAT 0209045;%%
Nucl. Phys. {\bf B} (Proc. Suppl.) {\bf 129} (2004) 263.

\bibitem{BMMRT} 
M. Bochicchio, L. Maiani, G. Martinelli, G.C. Rossi and M. Testa,  
Nucl. Phys. {\bf B262} (1985) 331.

\bibitem{KSUW}
K. Jansen, A. Shindler, C. Urbach and I. Wetzorke
[$\chi$LF Collaboration], Phys. Lett. {\bf B586} (2004) 432.
        
\bibitem{XLF}
W. Bietenholz {\it et al.} [$\chi$LF Collaboration],
JHEP {\bf 0412} (2004) 044, hep-lat/0411001 and 
Nucl. Phys. {\bf B} (Proc.Suppl.) {\bf 140} (2005) 683, hep-lat/0409109;\\
A. Shindler, K. Jansen, C. Urbach and I. Wetzorke, 
Nucl. Phys. {\bf B} (Proc.Suppl.) {\bf 140} (2005) 746;\\
K. Jansen, M. Papinutto, A. Shindler, C. Urbach, I. Wetzorke,
[$\chi$LF Collaboration], hep-lat/0507010

\bibitem{CAN}
A.M. Abdel-Rehim, R. Lewis and R.M. Woloshyn, Phys. Rev. {\bf D71}
(2005) 094505, hep-lat/0503007.

\bibitem{SWNEW}
S.R. Sharpe and J.M.S. Wu, Phys. Rev. {\bf D71} (2005) 074501, hep-lat/0411021.

\bibitem{AB}
S. Aoki and O. B\"ar, Phys. Rev. {\bf D70} (2004) 116011, hep-lat/0409006.

\bibitem{FMPR}
R. Frezzotti, G. Martinelli, M. Papinutto and G.C. Rossi, hep-lat/0503034.

\bibitem{PJ}
K. Jansen {\it et al.} [$\chi$LF Collaboration], hep-lat/0503031.

\bibitem{AO}
S. Aoki, Phys. Rev. {\bf D30} (1984) 2653 and 
%%CITATION = PHRVA,D30,2653;%%
Phys. Rev. Lett. {\bf 57} (1986) 3136.
%%CITATION = PRLTA,57,3136;%%
%
\bibitem{SHASIN}
S.R. Sharpe and R. Singleton, Jr.,
Phys. Rev. {\bf D58} (1998) 074501; hep-lat/9804028;\\
%%CITATION = HEP-LAT 9804028;%%
%
S.R. Sharpe and J.M.S. Wu, Phys. Rev. {\bf D70} (2004) 094029,
hep-lat/0407025.

\bibitem{MUSCO}
G. M\"unster, JHEP {\bf 0409} (2004) 035, hep-lat/0407006;\\
L. Scorzato, Eur. Phys. J. {\bf C37} (2004) 445, hep-lat/0407023.

\bibitem{KETAL}
F. Farchioni, R. Frezzotti, K. Jansen, I. Montvay, G.C. Rossi, E. Scholz,
A. Shindler, N. Ukita, C. Urbach, I. Werzorke,  
Eur. Phys. J. {\bf C39} (2005) 421, hep-lat/0406039
and Nucl. Phys. {\bf B} (Proc.Suppl.) {\bf 140} (2005) 240, hep-lat/0409098.

\bibitem{MONT}
F. Farchioni, K. Jansen, I. Montvay, E. Scholz, L. Scorzato, A. Shindler,
N. Ukita, C. Urbach and I. Wetzorke, hep-lat/0410031.

\bibitem{DWB2}
T. Takaishi, Phys. Rev. {\bf D54} (1996) 1050;\\
P. de Forcrand {\it et al.} [QCD-TARO Collaboration], Nucl. Phys. {\bf B}
Proc. Suppl. {\bf 53} (1997) 938; hep-lat/9608094.

\bibitem{SCH}
M. L\"uscher, R. Narayanan, P. Weisz and U. Wolff, Nucl. Phys. 
{\bf B384} (1992) 168;\\
M. L\"uscher, R. Sommer, P. Weisz and U. Wolff, Nucl. Phys. 
{\bf B389} (1993) 247 and {\it ibidem} {\bf B413} (1994) 481.

\bibitem{SINT}
S. Sint, Nucl. Phys. {\bf B421} (1994) 135 
and {\it ibidem} {\bf B451} (1995) 416.

\bibitem{QMR}
  S. Capitani, M. Luscher, R. Sommer and H. Wittig [ALPHA Collaboration],
  %``Non-perturbative quark mass renormalization in quenched lattice QCD,''
  Nucl. Phys. {\bf B544} (1999) 669, hep-lat/9810063;\\
  %%CITATION = HEP-LAT 9810063;%%
  J. Garden, J. Heitger, R. Sommer and H. Wittig  [ALPHA Collaboration],
  %``Precision computation of the strange quark's mass in quenched QCD,''
  Nucl. Phys. {\bf B571} (2000) 237, hep-lat/9906013;\\
  %%CITATION = HEP-LAT 9906013;%%
  J. Rolf and S. Sint [ALPHA Collaboration],
  %``A precise determination of the charm quark's mass in quenched QCD,''
  JHEP {\bf 0212} (2002) 007, hep-ph/0209255; \\
  %%CITATION = HEP-PH 0209255;%%
  M.~Della Morte, R.~Frezzotti, J.~Heitger, J.~Rolf, R.~Sommer 
  and U.~Wolff [ALPHA Collaboration],
%``Computation of the strong coupling in QCD with two dynamical flavours,''
  Nucl.\ Phys.\ B {\bf 713} (2005) 378, hep-lat/0411025.

\bibitem{RC}
   A. Bucarelli, F. Palombi, R. Petronzio and A. Shindler,
  %``Moments of parton evolution probabilities on the lattice within the
  %Schroedinger functional scheme,''
  Nucl. Phys. {\bf B552} (1999) 379, hep-lat/9808005;\\
  %%CITATION = HEP-LAT 9808005;%%
  M. Guagnelli, K. Jansen, F. Palombi, R. Petronzio, A. Shindler and I. Wetzorke
                  [Zeuthen-Rome (ZeRo) Collaboration],
  %``Non-perturbative pion matrix element of a twist-2 operator from the
  %lattice,''
  Eur. Phys. J. {\bf C40} (2005) 69, hep-lat/0405027;\\
  %%CITATION = HEP-LAT 0405027;%%
   M. Guagnelli, J. Heitger, C. Pena, S. Sint and A. Vladikas [ALPHA
                  Collaboration],
  %``Non-perturbative renormalization of left-left four-fermion operators in
  %quenched lattice QCD,''
  hep-lat/0505002;\\
  %%CITATION = HEP-LAT 0505002;%%
  M. Della Morte, R. Hoffmann, F. Knechtli, R. Sommer and U. Wolff,
  %``Non-perturbative renormalization of the axial current with dynamical Wilson
  %fermions,''
  JHEP {\bf 0507} (2005) 007, hep-lat/0505026.
  %%CITATION = HEP-LAT 0505026;%%  

\bibitem{LSSW}
M. L\"uscher, S. Sint, R. Sommer and P. Weisz, Nucl. Phys. 
{\bf B478} (1996) 365, hep-lat/9605038.

\bibitem{PT}
R. Narayanan and U. Wolff, Nucl. Phys. {\bf B444} (1995) 425;\\
S. Sint and R. Sommer, Nucl. Phys. {\bf B465} (1996) 71;\\
  A. Bode, U. Wolff and P. Weisz  [ALPHA Collaboration],
  %``Two-loop computation of the Schroedinger functional in pure SU(3)  lattice
  %gauge theory,''
  Nucl. Phys. {\bf B540} (1999) 491, hep-lat/9809175;\\
  %%CITATION = HEP-LAT 9809175;%%
  A. Bode, P. Weisz and U. Wolff  [ALPHA collaboration],
  %``Two loop computation of the Schroedinger functional in lattice QCD,''
  Nucl. Phys. {\bf B576} (2000) 517, hep-lat/9911018;\\
  %%CITATION = HEP-LAT 9911018;%%
  S. Sint and P. Weisz  [ALPHA collaboration],
  %``The running quark mass in the SF scheme and its two-loop anomalous
  %dimension,''
  Nucl. Phys. {\bf B545} (1999) 529, hep-lat/9808013.
  %%CITATION = HEP-LAT 9808013;%%


\bibitem{ML}
M. L\"uscher, Comm. Math. Phys. {\bf 54} (1977) 283.

\bibitem{PREV}
G.C. Rossi and M. Testa, Nucl. Phys. {\bf B163} (1980) 109; 
{\it ibidem} {\bf B176} (1980) 477 and {\it ibidem} {\bf B237} (1984) 442;\\
K. Symanzik, Nucl. Phys. {\bf B190} (1981) 1;\\
G. Marchesini and E. Onofri, Nuovo Cim. {\bf A65} (1981) 298:\\
G.C. Rossi and K. Yoshida, Nuovo Cim. {\bf 11D} (1989) 101.

\bibitem{TAN}
Y. Taniguchi, hep-lat/0412024.

\bibitem{SINTVM}
S. Sint, talk given at the ``Twisted Mass Lattice Fermions'' 
workshop held in Villa Mondragone (Frascati - Italy), March 14-15, 2005.

\bibitem{SYM}
K. Symanzik, in ``{\it New Developments in Gauge Theories}'',
page 313, Eds. G. `t~Hooft {\it et al.}, Plenum (New York, 1980);\\
K. Symanzik, ``Some topics in quantum field theory'' in
``{\it Mathematical Problems in Theoretical Physics}'',  Eds. R.
Schrader {\it et al.}, Lectures Notes in Physics, Vol. 153, Springer (New
York, 1982);\\K. Symanzik, Nucl. Phys. {\bf B226}  (1983) 187 and 205. 

\bibitem{MON}
I. Montvay and G. M\"unster, ``{\it Quantum Fields on a Lattice}'', 
Cambridge University Press (Cambridge, 1984). 

\end{thebibliography}
\end{document}